\documentclass[AMA,Times1COL]{WileyNJDv5} %STIX1COL,STIX2COL,STIXSMALL

\usepackage{lmodern}      % Latin Modern
\usepackage{amsmath,amssymb}

\articletype{Research Article}%

\received{Date Month Year}
\revised{Date Month Year}
\accepted{Date Month Year}
\journal{Journal}
\volume{00}
\copyyear{2026}
\startpage{1}

\begin{document}

\title{Thermoelastic wave-based logic for mechanically cognitive materials}

\author[1]{Ethan Fort}

\author[1]{Mohamed Mousa}

\author[1,2]{Mostafa Nouh}

\authormark{Fort \textsc{et al.}}
\titlemark{Thermoelastic wave-based logic for mechanically cognitive materials}

\address[1]{\orgdiv{Department of Mechanical and Aerospace Engineering}, \orgname{University at Buffalo (SUNY)}, \orgaddress{\state{Buffalo, NY 14260-4400}, \country{USA}}}

\address[2]{\orgdiv{Department of Civil, Structural and Environmental Engineering}, \orgname{University at Buffalo (SUNY)}, \orgaddress{\state{Buffalo, NY 14260-4300}, \country{USA}}}

\corres{Email: \url{mnouh@buffalo.edu}}

\abstract[Abstract]{Recent advances in metamaterials and fabrication techniques have revived interest in mechanical computing. Contrary to techniques relying on static deformations of buckling beams or origami-based lattices, the integration of wave scattering and mechanical memory presents a promising path toward efficient, low-latency elastoacoustic computing. This work introduces a novel class of multifunctional mechanical computing circuits that leverage the rich dynamics of phononic and locally resonant materials. These circuits incorporate memory-integrated components, realized here via metamaterial cells infused with shape memory alloys which recall stored elastic profiles and trigger specific actions upon thermal activation. A critical advantage of this realization is its synergistic interaction with incident vibroacoustic loads and the inherited high speed of waves, as opposed to recent adaptations of mechanically intelligent systems that employ innately slower mechanisms such as elastomeric shape changes and snap-through bistabilities. Through a proof-of-concept physical implementation, the efficacy and reconfigurability of the wave-based gates are demonstrated via output probes and measured wavefields. Furthermore, the modular design of the fundamental gates can be used as building blocks to construct complex combinational logic circuits, paving the way for sequential logic in wave-based analog computing systems.}

% No need for changes, left general w.r.t experimental validation

\keywords{mechanical computing, logic gates, smart materials, acoustic waves, memory}

\maketitle

\renewcommand\thefootnote{}

\renewcommand\thefootnote{\fnsymbol{footnote}}
\setcounter{footnote}{1}

\section{Introduction}\label{sec1}

In modern times, computing has largely been achieved using electronics. The fundamentals lie in Boolean logic\cite{boole1854}, which provides a universal framework for information processing, allowing computational architectures to be realized in any medium capable of distinguishing bits. Mechanical computing\cite{yasuda2021mechanical} offers one viable approach and has an established history of applications in tracking astronomical events\cite{freeth2006decoding}, solving differential equations\cite{bush1931differential}, and ballistics targeting\cite{mindell1995}. These analog computers benefited from real-time processing, energy efficiency, and task specificity, but fell out of favor following the arrival of digital systems, distinguished by fast pace, reprogrammability, and scalable designs. Despite this, the rapid developments in the space of adaptive and functional materials, combined with advancements in high-precision manufacturing and selective placement of properties in intricate metamaterial arrays, have given rise to the notion of mechano-intelligence, where information processing can be perceived as an inherent material property\cite{alu2025roadmap}.

Digital computational systems often reside within information-rich environments, subject to acoustic, optical, or thermal loads. Mechanical computers provide an ideal platform to directly interact with these stimuli in their raw (mechanical) form, foregoing the need for transduction and thereby conserving energy and reserving digital processing power for critical tasks\cite{hughes2019wave}. Recent mechanical computing research has leveraged intelligent materials\cite{mcevoy2015materials}, such as those utilizing somatosensitive soft materials to create robotic grippers\cite{truby2018soft} or those employing shape memory alloys (SMAs) to develop bio-inspired robots\cite{xu2024bioinspired,wang2024bioinspired}. In resonant systems, particularly, where passive circuit components surpass their acoustic counterparts in terms of dissipation and energy losses by orders of magnitude, mechanical computing provides an underexplored path to low-power computing, especially where such impinging loads, carrying the computational input, contain infinitesimal power such as sound pressure or minimal vibrations\cite{dubvcek2024sensor,masmanidis2007multifunctional}. As such, wave-based mechanical computing has recently gained traction, giving impetus to systems which exploit dispersion and guided wave scattering to conduct computational tasks, ranging from integro-differential operations\cite{mousa2024parallel} and to neuromorphic networks in elastic\cite{moghaddaszadeh2024mechanical,mousa2025dual} and optical\cite{shabanpour2024multifunctional,wu2019neuromorphic} domains. While the aforementioned systems successfully demonstrate intelligence, their computations are typically constrained to variations of the specific tasks they are designed to perform. In other words, they are not readily generalizable to perform a multitude of tasks in the way computers that harness Boolean logic and bit abstraction can. A large catalog of mechanisms for mechanical combinational logic has been demonstrated, including bistable structures\cite{song2019additively,lin2024digital,yang2025bistable}, soft conductive polymers\cite{el2022mechanical}, kinematically bifurcated modules\cite{xi2025kinematically}, multistable structures capable of ternary logic\cite{zhang2021hierarchical}, and origami-inspired\cite{treml2018origami} structures, with related designs also extending to sequential logic\cite{el2024intelligent}. While achieving combinational logic, many of these systems are static or quasi-static in nature, rely on the precise spatial or temporal application of mechanical force, moving parts, and suffer from slower computational speeds. Addressing these challenges has motivated the exploration of computing through wave propagation, whose manipulation is a defining capability of acoustic metamaterials.

Acoustic metamaterials represent a class of structures engineered from self-repeating building blocks, which collectively possess unique and otherwise unattainable properties, stemming from their optimized arrangement rather than the properties of the base materials\cite{cummer2016controlling}. Examples include bandgaps, i.e., frequency regions of forbidden wave propagation\cite{liu2000locally,al2018experimental}, filtering\cite{christensen2010all}, cloaking\cite{zigoneanu2014three}, wavefront steering\cite{li2013reflected}, and nonreciprocal transmission\cite{attarzadeh2020NR}. The ability of these artificially engineered materials to manipulate wave propagation provides a natural platform for information processing. A growing body of work uses metamaterials to perform wave-based logic, such as those exploiting wave interference in phononic crystals\cite{bringuier2011phase}, nonlinear effects in granular chains\cite{li2014granular} and nanomechanical resonators\cite{romero2024acoustically}, graded phononic crystals\cite{zhang2025programmable}, magnetically tunable ring resonators\cite{omrani2025magnetically}, and topology-optimized waveguides\cite{jafari2026passive}. In some of these systems, the logic gate is often fixed at the design stage or relies on multiple inputs of varying frequency. While the dispersive properties of metamaterials are typically fixed unless reconfigured, there is a body of work combining such systems with materials that exhibit dynamic properties\cite{brinson1993one,clark1993high,white2015programmable} to create tunable, smart, and stimulus-responsive metamaterials\cite{pierce2020adaptive,hu20203d,moghaddaszadeh2023local}. In this way, these reconfigurable metamaterials have an inherent memory of their states, which can be retrieved on demand by the controlled application of an external stimulus (e.g., a magnetically bistable phononic device\cite{bilal2017bistable}).

In this work, we capitalize on the capabilities of these reconfigurable metamaterials, as well as the rich features associated with wave propagation, to create mechanical structures that provide a scalable and modular. We present wave-based logic gates via a thermally-tunable locally resonant metamaterial network embedded within input and output paths that represent binary digital bits. This metamaterial network is combined with shape memory alloys, which, under small strains, undergo purely temperature-dependent phase transformations between the soft martensite and hard austenite phases. These transformations alter the unit cell's dispersive properties, enabling the system's operating frequency to be moved between energy-admitting and prohibitive states. This approach enables the creation of mechanical logic gates by controlling the bit states of inputs and outputs through one-dimensional arrays of metamaterial unit cells whose states are determined by reconfigurable sensors. The gate's operating frequency and identical geometric layout allow different logical operations to be fully compatible, enabling a scalable and modular design approach for complex networks. We validate the numerical concept with experiments, establishing a physical realization of wave-based logic for single operations, demonstrating their respective truth tables and ease of reconfiguration. We also analyze their computational speed through the comparison of different heating and cooling schemes. Through the combined theoretical-experimental framework presented here, we take a step closer to general-purpose mechanical computing by demonstrating complex combinational logic, a first step towards sequential logic in wave-based computers, and broader physical implementation.

\section{Wave-based Mechanical Logic}\label{sec2}

\begin{figure*}[ht]
\centerline{\includegraphics[width=\textwidth]{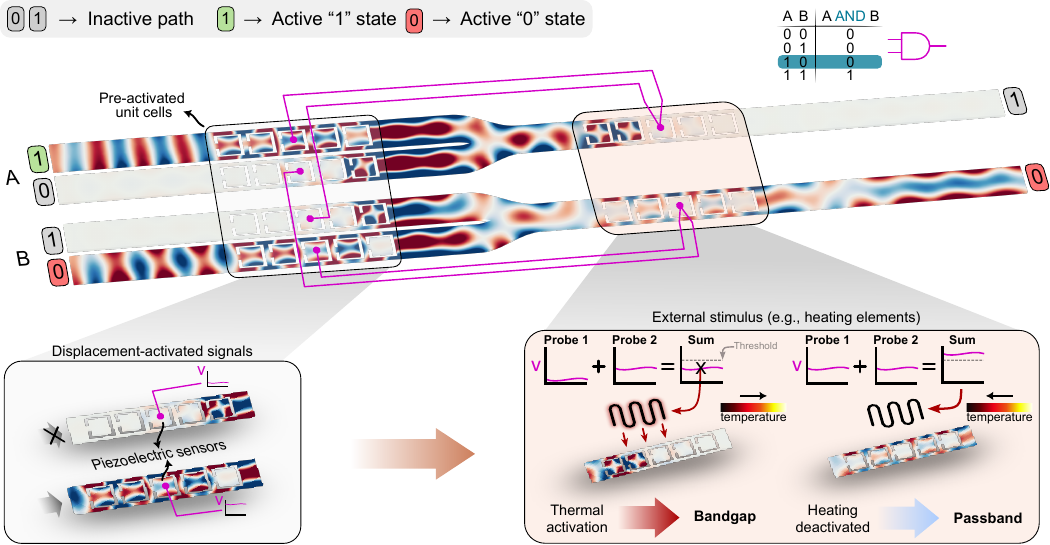}}
\caption{Conceptual design of a wave-based mechanical logic gate. Tunable elastic metamaterials in the boxed regions of the top figure drive the input-output relationship by controlling the flow of vibrational energy along the paths they are embedded within. An external stimulus (e.g., heating) allows the unit cells to retrieve passband or bandgap operational states to admit or block incoming vibrations, as shown in the bottom right schematic. Computational inputs marked by the excitations $A$ and $B$ propagate through pre-activated unit cells (active input $\rightarrow$ passband, inactive input $\rightarrow$ bandgap). A probing network (purple lines) detects input displacements (wavefields, bottom left box) and the network-driven mechanism located at the output paths applies stimulus (e.g., heat) to route energy to a high or low state, yielding the corresponding logic output.\label{fig1}}
\end{figure*}

To realize wave-based mechanical logic, a structure capable of distributing vibrational energy into discrete output channels, mappable to binary ``$1$'' and ``$0$'' states, is required. Figure~\ref{fig1} depicts the structural layout of the computing system, which leverages different sets of locally resonant elastic metamaterials positioned at different locations between the input (left) and readout (right) planes. The resonators housed within the metamaterial cells are stimuli-responsive and change their properties in response to changing ambient conditions (e.g., thermal or electrical loads), providing a path to tunable dispersion diagrams where an initial set of bandgaps exhibit frequency shifts as a result of altered material properties. The system can be designed around an operating frequency where, in an unstimulated state, waves are admitted due to the presence of a propagating mode, but are rejected upon stimulation by a given bandgap. This establishes an intrinsic element of memory in the system as it recalls different wave propagation regimes. In this way, vibrational energy can be manipulated to act as a computational carrier in a mechanically-intelligent system. In the following discussion, we present a foundation for wave-based mechanical logic and the further use of these logical building blocks in more complex circuits.

\subsection{Operational Concept}\label{sec2.1}

A tunable unit cell embedded within a path-based network, depicted in the boxed regions of Figure~\ref{fig1}, is used to perform logical operations by measuring displacements at the upstream input legs (left-side boxed region), which open or close the downstream legs to vibrational waves propagating in the paths (right-side boxed region). Thus, input excitations are routed through the open legs, thereby mapping vibrational states to conventional binary digits. At the input stage (legs labeled with $A$ and $B$), the two identical structures each have excitation receptacles corresponding to a ``$1$'' (upper leg) or ``$0$'' (lower leg) input. It is important to emphasize that in this realization, an active state of ``$1$'' or ``$0$'' is denoted by the presence of vibrational energy, and the absence of waves in a leg represents an inactive path. This serves a similar purpose as the $4$-$20$ mA signal standard, where a $4$ mA current acts as a live zero to help distinguish between a valid ``$0$'' state and a fault condition such as an open circuit or loss of power. Furthermore, each structure allows for one active input, while the entire gate can only have one active output, barring transient effects. 

External stimulation from previous logical operations in a chain of gates pre-activates unit cells in the input stage legs. Arrays in active legs are set to the passband state to admit incoming vibrations, and those in inactive legs are in a bandgap state to reject any forward or back propagating waves. Probes (left-side purple circles), such as piezoelectric sensors, are placed at the center of each array for input sensing. Active legs will see high displacement amplitudes from the open propagation path (bottom left box, lower wavefield), and bi-directional blocking from the bandgap facilitates highly attenuated vibrations in the center of inactive leg metamaterial arrays, thereby isolating measurements from reflections (bottom left box, upper wavefield). The probe signals control actuators, such as heaters (bottom right box), to stimulate the output unit cells and drive them to one of the stored states. The schematic in Figure~\ref{fig1} depicts the probing network configuration necessary for an AND operation. To achieve any other logical operations, the network can be reconfigured to combine different probe signals and state-triggering thresholds for the output arrays, as detailed in Supplementary Note 1 (Figures~S1--S2) for the remaining gates. This approach allows different combinations of inputs to be repeatably directed towards discrete outputs, similar to digital logic operations.
 
Figure~\ref{fig1} shows a mechanical AND gate as a representative example. The resulting wavefield demonstrates the I/O relationship for a given set of inputs. Incident excitations within the gate's operating frequency are applied at the ``$1$'' and ``$0$'' channels of the two legs, $A$ and $B$, respectively. As waves propagate through the preset unit cells, probes measure the kinetic energy generated by the passage of waves through the input legs. For an AND operation, the threshold is set such that only when both ``$1$'' input legs have propagating waves does their sum trigger the deactivation of stimulus in the ``$1$'' output leg, subsequently activating that output state. A similar control scheme is used for the ``$0$'' output leg, with the threshold made lower so that if any of the ``$0$'' legs are active, it will be the lower output path that is guaranteed to be open. For this example, the input combination is such that energy will be directed through the lower port, resulting in the AND operation $1 \land 0 = 0$. Each gate uses the same geometric layout, so all binary operations (AND, OR, NAND, NOR, and XOR) can be achieved through precise tuning of the probing network, and the unary NOT operation is shown using a modified gate structure. The established architecture ensures compatibility between connections of all the different logical operations, as the output of one gate can be physically connected to the next's input, sensor readings in inactive metamaterial arrays are isolated from the surrounding structure as waves are rejected from either side, and each gate is driven at the same operational frequency.  

\subsection{Thermally-activated Unit Cells}\label{sec2.2}

\begin{figure*}[ht]
\centerline{\includegraphics[width=\textwidth]{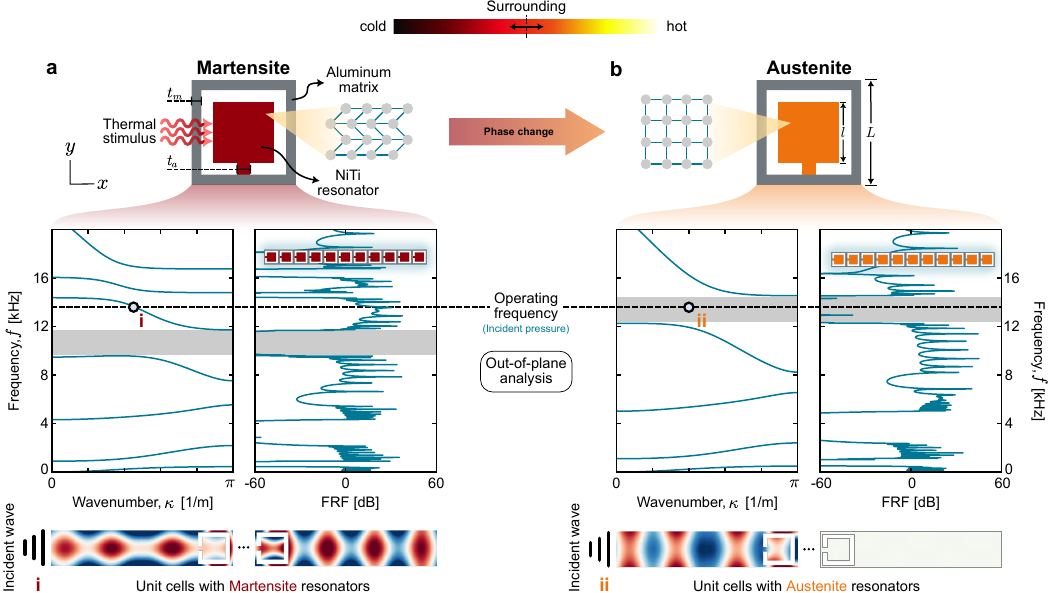}}
\caption{Tunable wave dispersion driven by temperature changes in the surrounding environment. (a) An elastic metamaterial unit cell with a shape memory alloy (SMA) local resonator under ambient conditions. The SMA resonator is comprised of a nitinol (NiTi) square mass connected to an outer aluminum frame via a slender NiTi neck. The unit cell is in a passband state (i) at $f_o = 13,650$ Hz, as shown in the dispersion diagram and frequency response plot. (b) Thermal stimulation reorients the internal lattice structure of the NiTi resonator to one with a higher elastic modulus, moving a bandgap around the operating frequency (ii). At the bottom of the figure, waveguide displacement fields compare the propagation of waves at the selected frequency under the two unit cell states, obtained via finite element modeling.\label{fig2}}
\end{figure*}

A prerequisite for wave-based logic using the current framework is a unit cell with programmable material properties, which are `remembered' depending on an applied stimulus. Here, we introduce a class of thermally active unit cells that change their stiffness, and consequently wave dispersive properties, in response to temperature changes\cite{candido2018tunable}. Shape memory alloys (SMAs) offer an interesting design avenue to reach this goal due to their unique thermomechanical behavior, and of particular interest, the ability to change their elastic modulus through a purely temperature-dependent phase transformation under low-stress conditions. Consider the locally resonant unit cell shown in Figure~\ref{fig2}, which is comprised of a square-shaped SMA resonator, specifically nitinol (NiTi), embedded within an outer matrix of aluminum with elastic modulus $E_{Al}=69$ GPa and density $\rho_{Al}=2,700$ kg/m$^{3}$. Nitinol is a common shape memory alloy with well-established material properties\cite{al2018experimental, liang1990constitutive}. The square internal mass has a length of $l=12.5$ mm and is connected to the outer ring by a neck of $t_a=2.5$ mm thickness. The outer ring has a side length of $L = 20$ mm and a thickness of $t_m = 1.25$ mm. At ambient temperatures, nitinol is in a softer martensitic phase, with an associated elastic modulus of $E_m = 26.3$ GPa and density $\rho_{n}=6,400$ kg/m$^{3}$, which generates the displayed out-of-plane dispersion analysis and frequency response, shown in Figure~\ref{fig2}a. Inspection of the dispersion curve reveals a bandgap (shaded region) from about $9.5$ to $12$ kHz, which is confirmed by the frequency response (i.e., transmissibility) plot. Sufficient thermal stimulation (typically $\sim 45~^\circ$C) lets the internal structure of the nitinol reorient itself into the higher stiffness austenite phase ($E_a = 67$ GPa, the same density is assumed) and consequently shifts existing modes into higher frequency bands. This is illustrated in Figure~\ref{fig2}b, where the denoted bandgap from the martensite unit cell has moved upward just under $3$ kHz and now spans the window from $12$ to $14.5$ kHz (shaded region). It is clear that at $13,650$ Hz, the martensite unit cell has a propagating path (i), but unit cells in the austenite phase reject the same excitation due to the stiffness modulation (ii), making this an ideal operational frequency, $f_o$. Thus, the computational memory is recalled via surrounding heat, triggering a future decision to impede or admit the flow of energy in the system, as can be seen in the pair of displacement fields of the waveguides shown at the bottom of the figure.

With the system equipped with the tunable unit cells necessary for the conceptual vision of a wave-based mechanical logic gate laid out in Section \ref{sec2.1}, going back to the AND gate example, the SMA-based unit cell can now be deployed along the input and output paths. The unit cells in the active input legs would be pre-activated with the elastic modulus of the martensite phase (passband), $E_m$, and those in inactive input legs set to the austenitic elastic modulus (bandgap), $E_a$. The probing network remains unchanged from Figure~\ref{fig1}, and as a result, the elastic modulus of the unit cells stationed along the ``$1$'' output changes to $E_a$ whereas that of the unit cells in the ``$0$'' output path changes to $E_m$, marking the effect of the SMA phase transformation and yielding the expected output, as confirmed by the propagating vibrational energy along the bottom leg. The system performs equally well for the different scenarios within the AND truth table, as well as across the remaining logic operations, with all such results tracked in Figures~S3--S8 of Supplementary Note 2.

\section{From Building Blocks to Integrated Circuits}\label{sec3}
\subsection{Full Adder}\label{sec3.1}
\begin{figure*}
\centerline{\includegraphics[width=\textwidth]{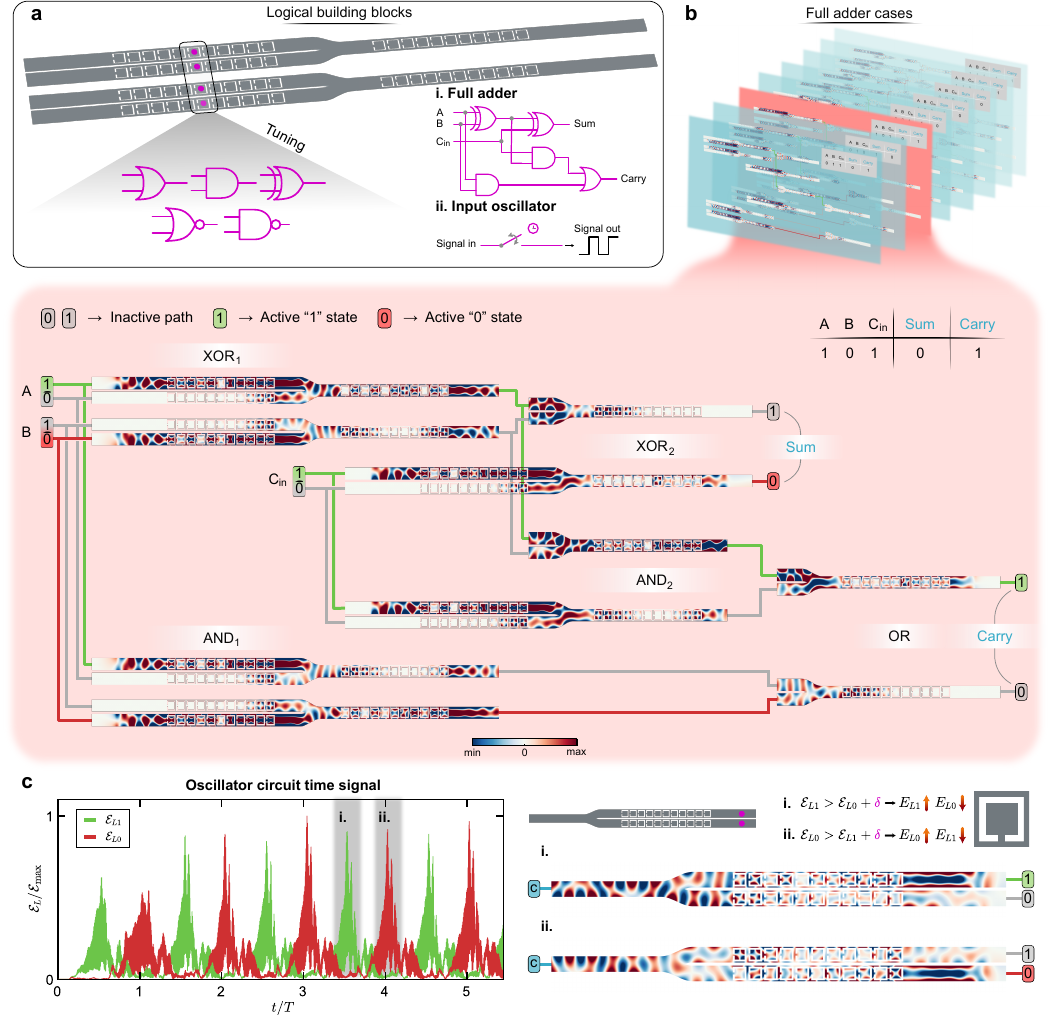}}
\caption{Mechanical logic gates for complex computational circuits. (a) The tuned sensor network (boxed region) enables any two-port mechanical gate to perform the various logic operations, such as OR, AND, XOR, NOR, and NAND. The gates function as building blocks for complex combinational logic circuits, such as a full adder (i), or, using a modified gate structure, an oscillator circuit, a critical prerequisite for sequential logic (ii). (b) Depiction of the eight possible full adder scenarios, with a detailed outset of one case shown in the light pink box: The wave-based mechanical adder receives vibrational inputs $A$, $B$, and $C_\text{in}$ on the far left side, prompting the two outputs $Sum$ and $Carry$ on the far right. The table in the upper right corner lists the binary I/O values for this case. Green, red, and gray lines represent active ``$1$'' and ``$0$'' paths, and inactive paths, respectively, while also showing the physical gate connections. (c) Mechanical input oscillator for clock cycle generation, the left plot demonstrates the oscillating behavior of the gate over a normalized time scale. The green and red signals are the normalized kinetic energy density within the upper ($\mathcal{E}_{L1}$) and lower ($\mathcal{E}_{L0}$) legs, respectively. The wavefields on the right show the displacement at the cycle peaks (i) and (ii). The circuit receives a constant excitation, and the expressions adjacent to the schematic (upper right) dictate the opening and closing of the output paths.\label{fig3}}
\end{figure*}

Having established a robust computing architecture capable of executing basic logic in mechanical form, demonstrated by the retrieval of wave-dispersive states through a tunable unit cell, we now have access to a toolbox of structures (See Figure~\ref{fig3}a) that can be utilized to integrate more complex mechanical circuits. To showcase the computational ability as well as the scalability and connectivity of the gates, consider a mechanical analog to a digital full adder, a foundational circuit in digital computing. As a reference, the schematic in Figure~\ref{fig3}a (i) shows the digital representation of a full adder, along with the labels for the inputs and outputs. The adder takes two $1$-bit inputs, the operands $A$, $B$, and another bit $C_\text{in}$, which represents the carry-in digit necessary for a sequence of full adders. The addition yields two outputs, $Sum$ and $Carry$. The $Sum$ corresponds to the addition result at the current binary place value, while the $Carry$ is the overflow digit propagated into the next place value. The full adder consists of five gates, two XOR, two AND, and a single OR. A full adder, therefore, has eight possible input combinations, resulting in a truth table with as many rows, as illustrated in Figure~\ref{fig3}b, where each slice represents one of these cases. For analysis, consider the input case $A=1$, $B=0$, and $C_\text{in}=1$, shown in the pink slice. The inputs and outputs are summarized in the table at the top of the pink box, and the wavefield resulting from the computation is shown in the center. 

The adder computes sequentially, meaning inputs at the earliest layers are computed first, and the resulting outputs are used in downstream gates. The first layer consists of the XOR$_{1}$ and AND$_{1}$ gates, which receive the $A$ and $B$ inputs, where green and red lines represent active ``$1$'' and ``$0$'' paths, and gray lines denote inactive paths. Excitations at operating frequency, $f_o$, are applied as displacements on vertical boundaries. Probes at the center of the XOR$_{1}$ gate's input array, consisting of eleven unit cells, detect that the ``$1$'' and ``$0$'' paths are active, and divert the energy through the ``$1$'' output leg according to the mechanical XOR's probing network.
    
Since the output of any gate becomes the inputs to subsequent gates, instead of having another (redundant) metamaterial array following the output path of a gate, we model gates in downstream layers as just half of the full gate (e.g., the upper port of AND$_{2}$). Following this, to read the result of computations, probes for the gates of the subsequent layers are located on the output resonators of the gates they connect with. Thus, the result from the XOR$_{1}$ gate is passed to the second layer of gates, which becomes one input of the XOR$_{2}$ and AND$_{2}$ gates, with the other input of those gates being the excitation from the carry-in bit ($C_\text{in}$). The computation of the other initial-layer gate, AND$_{1}$, is passed to the lower port of the OR gate in the last layer, while its other input comes from the AND$_{2}$ gate. Since the two-dimensional modeling of the gates does not allow for the three-way connections in a full adder, gates are connected using continuity boundary conditions. The XOR$_{2}$ gate computes the first output, the $Sum$, which, for this example, is computed based on the two active ``$1$'' states it detects, resulting in a ``$0$'' bit being output. Wrapping up the computation, the OR gate uses the outcome of the AND$_{1}$ (``$0$'') and AND$_{2}$ (``$1$'') gates, and computes the OR operation $1 \lor 0 = 1$, signaling that there is overflow and a bit will be carried into another adder. Thus, the adder computes the binary addition of the two `1' binary bits, $A$ and $C_\text{in}$, resulting in the outputs reading out as $Sum=0$ and $Carry=1$. The remaining addition cases are documented in Figures~S9--S16 of Supplementary Note 3. 

This system works as a standalone addition module that can be connected in parallel with other adders by feeding the carry-out of the adder in the current binary place ($2^n$), as the carry-in to the next adder ($2^{n+1}$), thus achieving n-bit addition. With this gating system, more n-bit combinational logic circuits are possible, ranging from multiplier circuits and multiplexers to encoders and comparators. Of course, the scaling is not without limit, as the metamaterial arrays do not have perfect transmission, and a physical system will have material damping to consider. Future work could address this by incorporating a wave-based buffer circuit to restore signal integrity. 

\subsection{Input Oscillator}\label{sec3.2}

While combinational logic circuits demonstrate the capacity for mechanical logic, any large-scale computing system needs a reliable timing reference to allow logic operations to be executed synchronously. For example, a serial adder reduces the number of adders needed to compute n-bit addition to just one by performing bit-by-bit addition, reducing the footprint of large-scale circuits, a criteria which is increasingly important in mechanical systems. However, a clock and memory elements (e.g., shift registers and flip-flops) are needed to instruct the adder when to execute operations and to store previous states. 

As a first step toward wave-based sequential logic, we introduce an oscillator circuit that functions as a mechanical clock by alternating between active ``$1$'' and ``$0$'' states, schematically illustrated in Figure~\ref{fig3}a (ii). To ensure compatibility with the mechanical logic gates, the clock design uses a single input leg that splits into two paths corresponding to ``$1$'' and ``$0$'' which contain the memory-integrated metamaterial. When the output legs are in opposing states, the input load is directed to the open path when a constant source of excitations at $f_o$ is applied. Such a signal is demonstrated in Figure~\ref{fig3}c, where the green and red curves represent the energy in the ``$1$'' and ``$0$'' output channels, respectively, recorded by probes located at the purple circles in the adjacent schematic. The model continuously updates the elastic moduli of the upper and lower leg resonators, which essentially relies on two conditions, a growth term (i) and a decay term (ii), depicted in the expressions on the lower right-hand side of the figure. Consider, when the clock is in a high state, shown in the graph as the shaded region and the corresponding wavefield labeled (i). The growth term is active, since the energy in the upper leg, $\mathcal{E}_{L1}$, is greater than the lower, $\mathcal{E}_{L0}$, plus some offset value, $\delta$. At the trailing and leading edges of cycles, there is a possibility of switching instabilities as the energy signals converge to comparable values, which is prevented using the offset. When this condition is met, the model increases the elastic modulus of the resonators in the upper leg, thereby closing the upper leg path while concurrently opening the lower leg path. After a transient period, the upper and lower legs are in the bandgap and passband states, respectively, thus cycling the clock to the low state (Graph and wavefield (ii)). Now the decay term will be active, and the system continues to oscillate between the two states. 

To generate the oscillating signal, this circuit relies on the transient effects of heating and cooling to increase and decrease the elastic modulus of the resonators. To accurately model the vibrations in the system, a small time step is necessary, i.e., $t_{\text{step}} = (20f_o)^{-1}$. Thermal effects, however, are slower, and to model both, a very long simulation would be needed. To circumvent this, we define a set of ODEs that continuously calculate a state variable based on the aforementioned conditions, with more details outlined in Section \ref{methods}. This state variable is used to set the elastic modulus value of the upper and lower leg resonators to generate the oscillating response. Thus, by controlling the rate at which the state transforms with respect to the present output energy, a faster thermal response can be modeled. As such, the timescale has been normalized to the average period of the cycles ($T=5.7$ ms), to give a qualitative view of the oscillating nature of the circuit, rather than quantitative. The displacement field is animated to better depict the oscillating nature of the circuit in Supplementary Movie 1. 

\section{Physical Realization of Wave-based Mechanical Logic}\label{sec4}
\subsection{Experimental Setup}\label{sec4.1}

\begin{figure*}[ht]
\centerline{\includegraphics[width=\textwidth]{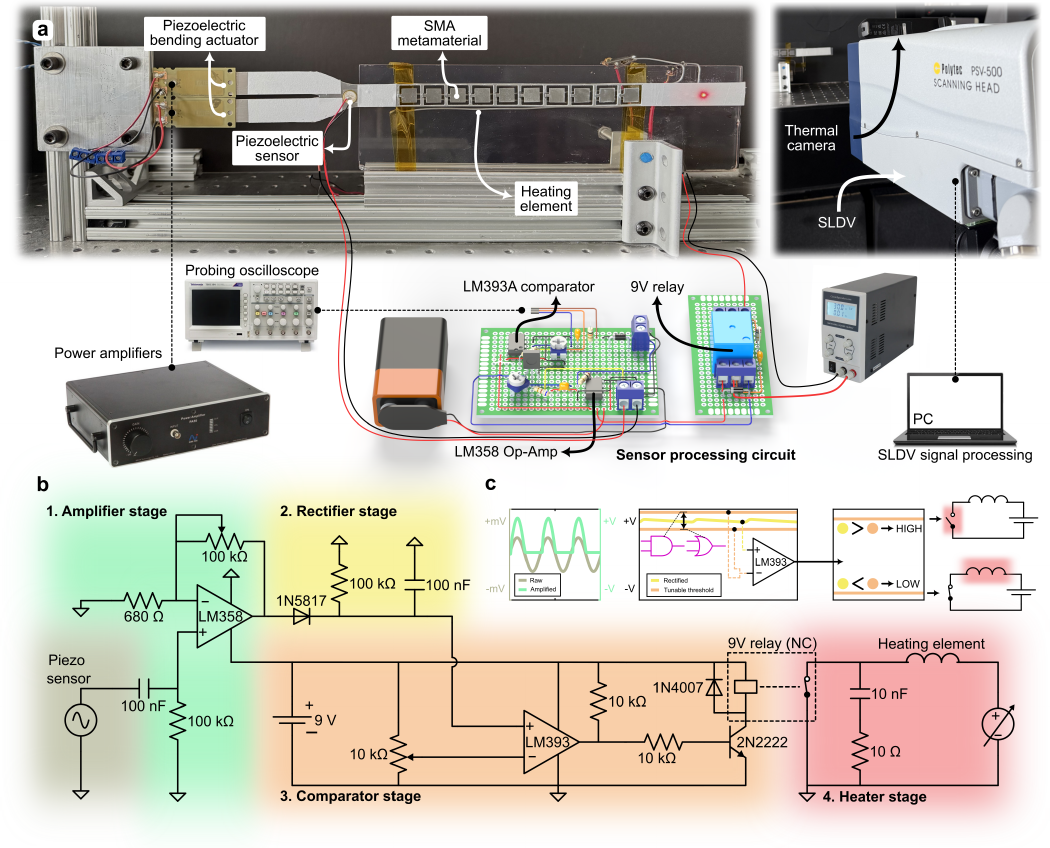}}
\caption{Physical realization of wave-based mechanical logic. (a) Photograph of the experimental setup. Upper images present the single-port design of the mechanical logic gate as well as the laser Doppler vibrometer scanning head and the thermal camera used to capture displacement and temperature field data, respectively. The middle section shows the equipment used to operate the gate. (b) Circuit schematic of the sensor processing circuit (rendering shown in the middle section). (c) Evolution of the piezoelectric sensor voltage through the processing circuit stages obtained from SPICE simulation. Details are provided in Section~\ref{sec4.1}.\label{fig4}}
\end{figure*}

In the previous sections, we charted a path to wave-based mechanical logic and demonstrated the integration of various gates within a complex network. Here, we present a viable experimental implementation of wave-based mechanical logic using the envisioned framework, a thermally-tunable structure capable of combinational logic. Unlike the simulated architecture, which defined both the ``$1$'' and ``$0$'' states as waves propagating through their respective paths, the experimental design adopts a single structure (photographed in Figure~\ref{fig4}a), for simplicity. As a result, the ``$0$'' state is henceforth defined by the absence of propagating waves, which in practical terms is denoted by propagation amplitudes which fall below a threshold. This is consistent with a $0-20$ mA architecture and reduces the overall footprint of the logic gates. This simplified single-port implementation provides a compact architecture for experimental validation, but introduces an additional consideration for physical cascading because multiple active inputs can contribute to the response of the same output path. Consequently, the amplitude representing a logical state can depend on the input combination. The dual-port architecture used in the numerical studies avoids this particular superposition-dependent amplitude variation by assigning the logical ``$0$'' and ``$1$'' states to separate propagation paths, at the cost of a larger physical footprint. Both architectures remain compatible with physical cascading, although their transmission and amplitude-management requirements differ, as discussed further in Supplementary Note 7.

While the conceptual paradigm used here is indifferent to the mode of excitation, we focus in these experiments on out-of-plane displacement modes to best visualize the wavefield using a 1D Scanning Laser Doppler Vibrometer, which measures motion normal to the field of vision. Piezoelectric bending actuators are fixed on the two input legs, the $A$ and $B$ inputs of the logic gate, to generate out-of-plane excitations. Two BSWA PA50 amplifiers independently power the benders, which allows unique gain tuning of each actuator to mitigate differences in their responses (Actuator tuning process is detailed in Supplementary Note 4 and Figures~S17--S18). All experiments are performed with the same amplifier gains to ensure consistency between tests. Furthermore, the metamaterial arrays in the input paths are neglected, similar to the gates in downstream layers in the full adder (See Figure~\ref{fig3}b), since the activation of either source implies that waves virtually passed through unit cells in a passband state. Recall that in the previous framework, an active ``$0$'' state was denoted by the existence of waves in the associated legs, whereas now, this state is active when energy is absent. Nitinol sourced from Nexmetal Corporation is used to fabricate the thermally-tunable metamaterial, which is martensitic under ambient temperatures and fully austenitic above $50~^\circ \textrm{C} \pm 10~^\circ \textrm{C}$, per their specifications. The natural frequencies of the nitinol sample under ambient and elevated temperature conditions were used to characterize the elastic modulus. The corresponding frequency sweep results are shown in Figures~S19 and~S20, and details of the test are documented in Supplementary Note 5. The martensite and austenite elastic moduli were found to be $E_{m,e}=56$ GPa and $E_{a,e}=82$ GPa, respectively. We use the unit cell design established in Section \ref{sec2.2}, except that the length and width dimensions are scaled down by $20\%$. The thicknesses of the aluminum and SMA are $0.5$ mm and $0.31$ mm, respectively. This scaled design allows for increased thermal performance in the experiments. Finally, metal-bonded epoxy is used to adhere the resonator to the top surface of the aluminum outer matrix.

To control the application of heat to the unit cells, a piezoelectric disc placed before the output path senses displacements from the actuators. Since the sensor generates a noisy low-voltage AC signal, it is processed via a custom sensor processing circuit, illustrated in the rendering below the gate. The circuit consists of four main stages (See Figure~\ref{fig4}b), and aided by SPICE simulation data in Figure~\ref{fig4}c, the evolution of the raw signal as it passes through the board can be tracked. Colored signals in the plots result from the similarly-colored regions of the circuit schematic. At the amplifier stage, an LM358 Op-Amp takes the small mV-scale AC voltage from the sensor (gray) and boosts it up to a V-scale signal (green). A variable potentiometer sets the amplification gain to prevent saturation when both input channels are active, allowing the use of the entire voltage range with different input combinations. A Schottky diode (1N5817) rectifies the amplified signal, and an RC filter applies further smoothing to create a clean DC signal (yellow). Using an LM393 comparator, the displacement-sensitive DC signal is compared to a potentiometer set threshold voltage (orange, second plot). The comparator generates a high or low voltage (orange, third plot), which toggles a normally closed 9V relay to turn off the heater when the input signal is greater than the reference; otherwise, the heater is left on (red). A flyback diode (1N4007) is placed across the relay contacts to protect the circuit from noise and voltage spikes, and an arc snubber circuit is added in parallel to the heater for further protection. 

With this design, simply tuning the comparator reference voltage allows the gate to be configured to perform different logic operations. With either input actuator on, the voltage generated by the piezoelectric sensor will be constant, and with both, the voltage effectively doubles. In other words, a reliable mapping between the active input legs and the rectified output voltage of the piezo processing circuit is achieved, which can be used for tuning the gate to perform different operations. For the experiments, full-field vibration measurements are captured using a Polytec PSV-500 scanning laser doppler vibrometer (SLDV), and unit cell temperature data is gathered with a FLIR ONE Edge thermal imaging camera. A Tektronix TBS1104 oscilloscope is used to monitor and record signals within the sensor processing circuit.

\subsection{Full-scale Testing}\label{sec4.2}

\begin{figure*}[h!]
\centerline{\includegraphics[width=\textwidth]{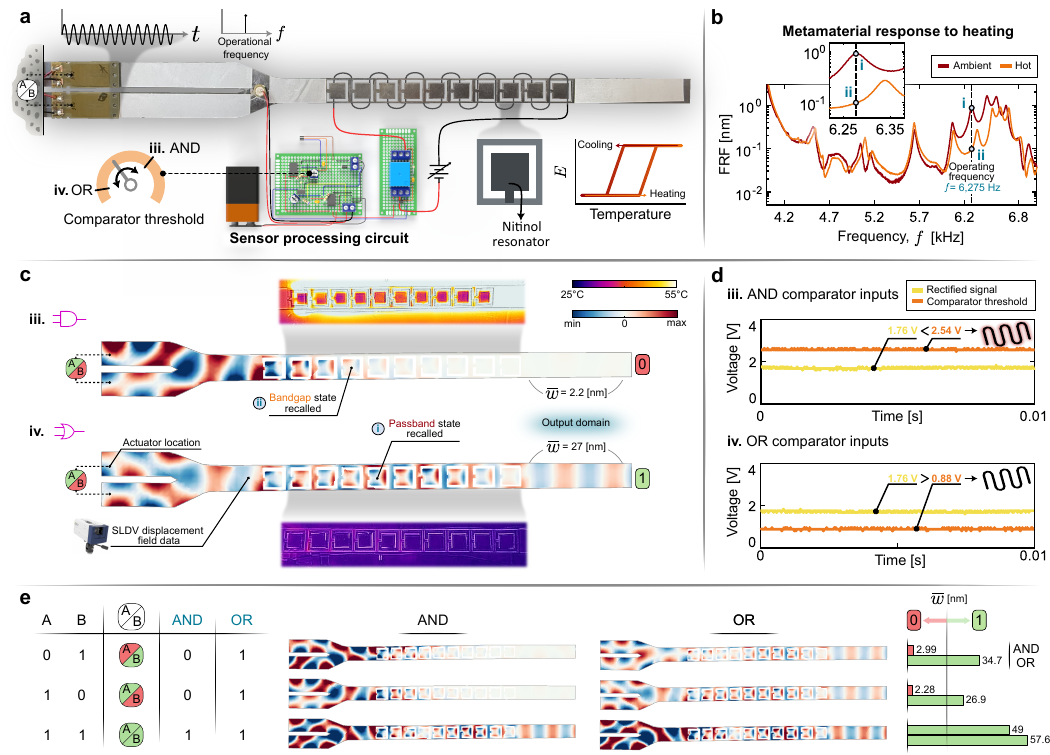}}
\caption{Experimental testing of the AND and OR operations. (a) In the upper left, inputs $A$ and $B$ are applied as continuous out-of-plane bending waves at the operating frequency. At the gate center, a piezoelectric sensor generates a voltage proportional to the number of active inputs. At the bottom left, the comparator threshold voltage configures the gate to perform AND operations at a higher value (iii) or OR operations at a lower value (iv). Behind the output path, the circuit activates or deactivates a heating element based on the comparator state. (b) Experimental frequency response plot of the dual-material unit cell at ambient (red) and hot (orange) temperatures. The inset depicts the frequency shift resulting from elevated temperatures. Recall, elevated temperatures cause the nitinol resonator to undergo the austenite phase change, raising the elastic modulus, and thereby shifting the frequency response rightward. The operating frequency, $f_{o,e}=6,275$ Hz, is chosen such that at ambient (i) and elevated (ii) temperatures, the system resorts to passband and bandgap states, respectively. (c) Displacement fields of the AND (iii) and OR (iv) logic operations with inputs $A=1$ and $B=0$. For the AND operation, elevated temperatures (upper thermal image) recall the metamaterial cells to a bandgap state (ii). For the OR operation, the same inputs deactivate the heater (lower thermal image), moving the system back to the passband state (i). (d) Oscilloscope data of the rectified sensor signal and comparator thresholds. Consequent heater state is shown for the AND (iii) and OR (ii) operations. (e) Truth table summary of AND and OR operations. The bar chart on the right compares the average displacement magnitude of the output domain for each case.\label{fig5}}
\end{figure*}

Experimental results for the AND and OR logic gates are shown in Figure~\ref{fig5}. Figure~\ref{fig5}b shows the transmissibility of the nitinol unit cell under ambient and heated thermal conditions. The logic gate could be subject to asymmetric loads, since it's possible for just one input leg to be excited at a time. Therefore, the sweep was performed with the top leg actuator ($A$) to activate the modes of the most inclusive excitation case. Inspection of the experimental frequency response under ambient conditions (red) indicates that there is a bandgap spanning the $4,500$ - $6,200$ Hz range. The nitinol phase transformation follows a hysteresis loop depending on whether it is heating or cooling, see the bottom right of Figure \ref{fig5}a. However, once fully martensitic or austenitic, it takes on a fixed value\cite{al2018experimental}. As such, we observe a rightward shift in the transmissibility curve upon applying external heating to the unit cells (orange). A detailed wave propagation analysis is provided in Supplementary Note 6, with numerical and experimental studies of the unit cell presented in Figures~S21 and~S22. An operating frequency of  $f_{o,e}=6,275$ Hz is chosen for the gate corresponding to a resonant mode at the right edge of the ambient bandgap (i), such that when heat is applied, the resulting bandgap shift drops the frequency into a region of high attenuation (ii). Further inspection suggests that other frequencies may work, so a detailed analysis of the operating frequency selection process is documented in Supplementary Note 7.

A concurrent analysis is presented of the AND and OR scenarios under identical input conditions to demonstrate the operation and reconfigurability of the mechanical logic gates. To configure the structure for AND operations, the comparator reference voltage is set to $2.54$ V. Channel $A$ is activated ($A=1$) and incident vibrations at the operating frequency begin propagating through the system, while channel $B$ is left dormant ($B=0$). The resulting displacement is received by the sensor, and the processing circuit outputs a constant $1.76$ V rectified signal, which is compared to the reference in Figure~\ref{fig5}d. The input voltage is not enough to open the relay, so the heating element remains on and raises the temperature of the unit cells. After a transient period, the temperature in the unit cells levels out to $\sim 45~^\circ$C, shown in the thermal snapshot above the gate, and at this point, the SLDV captures the displacement field of the gate. While this input is applied, the flow of energy to the output will continue to be blocked due to the thermal stimulus recalling the metamaterial into its bandgap state. Now consider an OR operation with the same inputs (See Figure~\ref{fig5}c iv) dictating an output of ``$1$''. To achieve this, the threshold voltage is simply lowered from $2.54$ V (AND) to $0.88$ V (OR), rendering the input voltage greater than the reference (See Figure~\ref{fig5}d iv), the heater is turned off, the gate cools to ambient conditions, and the system admits vibrational energy through the unit cells to the output. To quantify the gate's efficacy, the average displacement magnitude within the output domain, $\overline{w}$, is measured and found to be $2.2$ nm and $27$ nm for the AND and OR cases, respectively. This difference of nearly $12$ times between low and high states confirms the gate's ability to shut off or admit energy to the readout leg in a robust manner, depending on the intended operation. A joint truth table summarizes the AND and OR results in Figure~\ref{fig5}e using the experimental procedure established above. A bar chart on the right-hand side compares the magnitude of displacement across the different gates under similar input conditions. The complementary input cases (Row $1$ and Row $2$ of the truth table) result in similar high and low state displacement magnitudes, with the OR gate having $\overline{w} = 34.7$ nm and $\overline{w} = 26.9$ nm for the high state, while the AND gate has $\overline{w} = 2.99$ nm and $\overline{w} = 2.28$ nm for the low state, respectively. Although the experimental gates maintain a clear distinction between the logical ``$0$'' and ``$1$'' output states, attenuation through the fabricated metamaterial results in a reduction in the displacement amplitude associated with the initial ``$1$'' input state. Such attenuation presents an important consideration for physical gate cascading, as losses accumulated over successive gates can reduce the margin between logical states. A more detailed discussion of amplitude variation, transmission, physical gate connections, and such limitations is provided in Supplementary Note 7 and Figure~S23. For both gates under $A=1$ and $B=1$ loading, the strength of the high state is greater than that of the remaining two OR cases, which yield the same output of ``$1$'', since both inputs are active and push more vibrational energy to the output path. Note that the $A=0$ and $B=0$ case for both operations is not presented, as it is a trivial result (no vibrations in the system). Phase animations of the displacement fields for the discussed operations are provided in Supplementary Movie 2.

\subsection{Transient Performance}\label{sec4.3}

While most attempts to implement mechanical logic to date have squarely focused on transitions, shape changes, or reconfigurations of multistable materials, tapping into beam buckling\cite{song2019additively}, origami lattices\cite{yasuda2017origami}, or soft conductive elastomers\cite{el2021digital}, the static nature of such systems has generally resulted in an inherently slow response, and as a consequence, computational pace. While it is not the intent of mechanical computing to compete with its digital counterpart, operational speed remains critically important. Wave propagation as a computing medium provides a clear path towards computational speeds that are orders of magnitude faster than quasi-static deformations in non-dynamical systems. The forthcoming analysis in this section highlights two important pillars of our wave-based architecture. First, that the computational pace is set by the rate at which unit cells can be heated and cooled. Thus, by improving the heat transfer rates, drastic improvements can be made to the overall task duration. Second, that the theoretical ceiling for the speed of the different logic operations shown earlier is dictated by the waves' time of travel in the pertinent waveguides, if the aforementioned thermal transient effects can be minimized. In addition to analyzing gate performance, this experiment demonstrates the capability of the logic gate to respond, with no further adjustments, to changing input conditions, as evident by its ability to self-adapt and effectively transition from Task 1 to Task 2, signaling its readiness to adequately function in dynamic and rapidly changing environments. 

\begin{figure}[ht]
    \centering
\includegraphics[width=\textwidth]{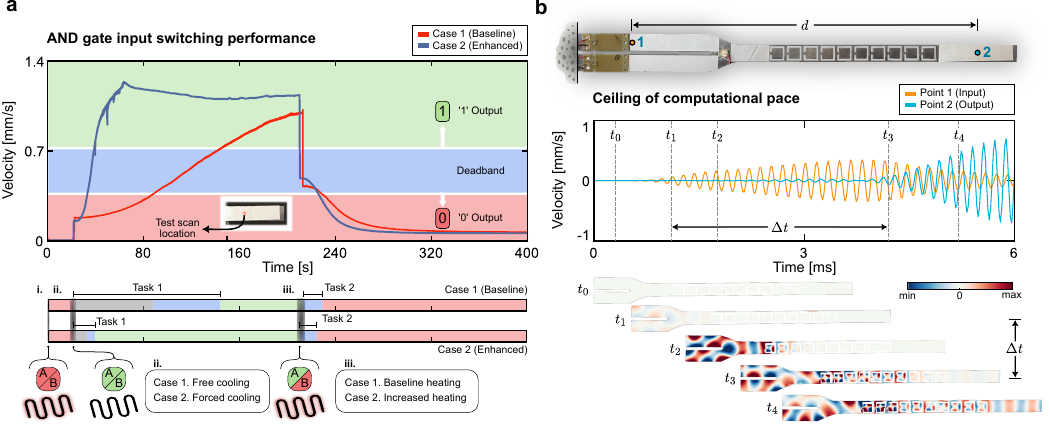}
\caption{Computational pace analysis. (a) Performance demonstration of the AND gate, considering transient thermal effects during passband and bandgap state recollection. The gate is initialized in a heated state (i), and the inputs are switched two times at (ii) and (iii), and the velocity signal is recorded. The switching is performed under both a baseline heating and cooling regime (red), and an enhanced regime (blue), which utilizes a computer fan and increased heating power to speed up computations. The output state is projected into timelines located at the bottom of the figure. (b) Analysis of the ceiling of the computational speed of wave-based logic in the absence of thermal delays. The plot on the right depicts the velocity signal at a point near the channel $A$ actuator (orange), and another point located at the output (light blue). A burst of waves at the operating frequency is released just after $t_0$, and travels a distance $d=30$ cm in $\Delta t=3.1$ ms. Wavefields at the bottom right depict the propagation of the waves as they move towards the output. \label{fig6}}
\end{figure}

We begin by examining the performance of the AND logic operation when subjected to changes in its inputs, shown in Figure~\ref{fig6}a. With the gate configured to perform AND operations, three input cases are tested in succession (``$00$'', ``$11$'', and ``$10$''), and a single point scanned at the output records the response as it changes over time. Three output regions are defined: ``$0$'' output (light red), ``$1$'' output (green), and a deadband (light blue), a region where the system is known to be transitioning. The boundaries of these regions are chosen to mitigate the effect of instantaneous signal jumps when the inputs are initially switched. To visualize the switching performance, the analog velocity signal (top left of the figure) is projected into a state timeline (bottom left of the figure). In the discretized timeline, a fourth state (gray) corresponding to errors is included, defined as intervals where the measured signal lies outside the correct logic state region and has not yet transitioned to the deadband. 

First, an analysis of the system in a baseline operational state, under free cooling and $5$ W heating, is provided (red time signal in Figure~\ref{fig6}). The baseline performance case was used for the experimental results in Section~\ref{sec4.2}. The gate is initialized with $A=0$ and $B=0$, and the heater is left on until the temperature stabilizes (bandgap active). At this point (i), the scan is started, and after a $20$-second hold period to verify there are no initial vibrations (ii), both actuators are powered ($A=1$ and $B=1$). Due to the imperfect blocking of the unit cells, some energy is picked up in the time signal (the initial jump). However, the input vibrations instantly deactivate the heater, and the signal begins to grow as the gate cools. After $88$ seconds, the signal enters the deadband, and at $143$ seconds, it crosses into the ``$1$'' state. At $210$ seconds (iii), one of the inputs is deactivated ($A=1$ and $B=0$), and the signal immediately drops to the deadband. Simultaneously, the sensor processing circuit recognizes the change and begins heating the unit cells, causing the signal to decay into the ``$0$'' output region (crossover at $230$ seconds). The same test is repeated for Case 2, but the system is now subject to forced cooling via a computer fan placed in front of the array, and the heater power is increased to $8$ W. For the first case, the total computation time from (ii) to the correct output (i.e., Task 1) was around $122$ seconds, but the same transition takes just $18$ seconds when the fan is introduced. For the transition from (iii) back down to the low state (i.e., Task 2), the baseline case reached the correct output in $16$ seconds, compared to $14$ seconds for the test with the heater power raised. 

This analysis shows that the computational speed is largely set by the rate at which we can drive the tunable unit cells to their stored dispersive states. The computations requiring cooling saw the biggest change, with an $85$\% reduction in computation time with an approximate doubling of the heat transfer coefficient from the baseline case (natural vs. forced convection). On the other hand, the $3$ W heating power increase from the baseline is much less, $60$\%, and as expected, the reduction in computation time is only $13$\%. While these were relatively simple changes to increase the performance of a bench-top prototype, they indicate that robust and more advanced cooling and heating schemes would improve these times further. Moreover, the architecture proposed is not limited to just thermally-tunable metamaterials; other unit cell designs could utilize different materials with faster responses to other external stimuli. Within the scope of this work, different shape memory alloy compositions with transition temperatures closer to ambient temperatures could drastically decrease computation times since propagation paths could be adjusted faster with greater efficiency. 

We have demonstrated that the flow of information through vibrations is controlled by the thermally driven tuning of the unit cells. In other words, the input waves have the potential to transmit output state information many orders of magnitude faster, but are limited by our ability to heat and cool the system. When taken to the extreme, if the unit cells could be heated and cooled instantaneously, the wave speed becomes the new ceiling on the computational pace. This provides a promising outlook for wave-based logic to perform operations at the order of hundreds of operations per second. This limitation is first revealed in Figure~\ref{fig6}a by the signal jumps when the inputs are altered at times (ii) and (iii). The change in input conditions propagates through the system almost instantly, but the correct output determination lags as the heating or cooling takes effect. 

We aim to evaluate the potential computational speed of this architecture in the absence of thermal effects by computing the speed of waves in the gate (under ambient conditions). A pulse of waves at the operating frequency is sent into the system, and by comparing the arrival time of the excitations at two points, one at the input and another at the output, the wave speed can be determined. The points are depicted at the top of Figure~\ref{fig6}b, and the distance between them is $30$ cm. The plot in the middle of the figure shows the velocity evolution of the points as the pulse propagates. Various times are marked, and their corresponding wavefields are shown at the bottom. At $t_o=0.3$ ms, the system is in a steady state, and shortly after (at $0.5$ ms), waves are released into the system. The excitation begins to propagate, and at $t_1=1.1$ ms, the velocity signal crosses a threshold of $50$ $\mu$m/s, which is used to define the arrival of waves at point $1$. An intermediate snapshot is shown at $t_2=1.76$ ms, showing that the excitation has grown stronger at point $1$ and the wavefront has reached the metamaterial array. At $t_3=4.2$ ms, the vibrations reach the output having crossed the $50$ $\mu$m/s threshold. Notice that the signal at point $1$ drops below the maximum, signaling that the pulse has passed. Finally, at $t_4=5.2$ ms, the pulse has fully developed through the system, and the remaining waves are solely from resulting reflections. The time of travel between the input and output is calculated from the difference between $t_3$ and $t_1$, resulting in $\Delta t=3.1$ ms. Supplementary Movie 3 provides an animation of the time-varying displacement field as the tone burst travels through the gate. Using the difference in the wavefront arrival times, the wave speed is estimated with the associated distance $d$, resulting in $c=105.7$ m/s. The flexural wave speed in the experimental logic gate is given by $c_f=(\frac{E_{Al}h^2}{12\rho_{Al}})^\frac{1}{4}\sqrt{2\pi f_{o,e}}$ from the Euler-Bernoulli beam theory\cite{giurgiutiu2014structural}. For comparison with our experimentally calculated wave speed, we assume the gate to be a homogeneous $h=0.5$ mm thick beam of aluminum (at ambient temperature). Thus, we get a theoretical wave speed of $c_f=96.8$ m/s, which is comparable to our experimentally determined value.

Under the enhanced performance regime, the total time to complete Task $1$ was $18$ seconds. If the thermal effects are neglected (i.e., at (ii) the system instantly transitions from a bandgap to a passband state), the wave speed would set the computation time. While $\Delta t$ would provide a rough estimation of the new computation time, the true time would likely be a multiple of this, since the velocity signal should rise over some threshold to be considered a `$1$' state. First, the threshold of the `$1$' state should be determined based on the velocity signal in Figure~\ref{fig6}b. The threshold in the thermal performance test should not be used since the velocity signals do not reach the same level, likely due to differences in the experimental method used for the tests (See Section~\ref{methods}). In the performance study, for the output to be in state `$1$', the signal was required to cross around $0.7$ mm/s. The maximum recorded velocity across both performance cases was around $1.2$ mm/s, thus the threshold for a `$1$' state was approximately $70$\% of the maximum. The maximum velocity recorded at the output for the time of travel experiment was $0.7$ mm/s, meaning that for a meaningful comparison, the velocity must rise over $0.5$ mm/s to be in the `$1$' state. This crossover occurs just after $t_4$, meaning the computation time would be $t_4-t_1=4.1$ ms or about $1.3\Delta t$. Comparing the task times, the computation would be carried out nearly $4,400$ times faster when neglecting the thermal effects. It should also be emphasized that this value could potentially increase by an order of magnitude if the logic gates exploited in-plane deformations due to the inherently faster speeds of longitudinal waves (out-of-plane modes were used here to allow for the wavefield to be captured using a single-head SLDV system) or with a different choice of operational frequency (due to their frequency-dependent dispersive nature). These factors reiterate that wave-based mechanical logic has the potential to be conducted at a millisecond scale, if highly-responsive materials are utilized in the tunable unit cells alongside accelerated heating and cooling techniques. 

\section{Conclusions}\label{sec5}
In summary, this research presents a versatile framework for wave-based mechanical logic. Through combined simulations and experiments, we demonstrated the ability to inherently process applied vibroacoustic inputs through an integrated sensing network capable of abstracting energy into discrete bit states via paths embedded with thermally-tunable elastic metamaterials. In the numerical studies, we introduced a shape memory alloy-based unit cell which recalls stored states based on displacement detection in the input legs of the mechanical logic gates, thereby simulating the effect of SMA phase change under heating. This enables paths in the logic gates embedded with the tunable unit cells to independently open or close themselves to the vibrations at the (single) operating frequency due to underlying passbands and bandgaps. In the experiment, we implement and validate the envisioned unit cell in a bench top setup. With this framework, we numerically demonstrate scalability from single logic operations to complex networks, open the door for sequential wave-based logic by introducing a mechanical clock, and experimentally demonstrate effective logic operations (AND, OR) enabled by an analog sensor network to control heat application to the gates. Transient thermal effects were also analyzed by comparing two heating and cooling regimes, which illustrated considerable improvements in computation speed with minor changes (forced cooling and increased heating element power), and highlighted the potential of the proposed framework, with the mitigation of such effects, to conduct operations at a notably high pace relative to analog mechanically intelligent systems. The present experiments were designed to validate individual logic-gate operation, while system-level cascading (i.e., a full adder) is demonstrated numerically. Experimental integration of multiple gates, along with relevant signal-restoration considerations, is therefore left for future work. Ultimately, this work represents a scalable, adaptable, and modular approach to mechanical logic. As long as each gate has its three constituent components, namely the paths, the metamaterial arrays, and the sensor network, operation should not be hindered. These can be viewed as a general architecture necessary for wave-based computing, which relies on switchable energy-admissive and energy-blocking states, but the specific functionality of each can be modified to fit different applications. Changing unit cell geometry can tune the system to different operational frequencies, which may contain critical information about the environment within which they reside, allowing direct computation and subsequent action to be taken based on the presence (or absence) of that information. Furthermore, without loss of generality, different tuning mechanisms can be employed based on the availability of external stimuli within the operating environment. Finally, given well-established analogies between optical\cite{wu2019neuromorphic} and elastoacoustic\cite{moghaddaszadeh2024mechanical} wave-based computing systems, the blueprint presented here could be readily adopted across different physical domains, bringing such systems a step closer to general-purpose and readily implementable realizations.

\section{Methods}\label{methods}

COMSOL Multiphysics 6.3 was used for all numerical simulations and presentation of wavefields. The plate physics module was utilized for all studies, assuming $1$ mm out-of-plane thickness. Dispersion diagrams for the SMA-based unit cells were solved using the eigenfrequency study, with periodic conditions on the upper and lower boundaries, with respect to the unit cell shown in Figure~\ref{fig2}, with only out-of-plane wave modes considered. Frequency response (transmissibility) plots were obtained from a finite structure ($11$ unit cells) subject to an edge load excitation over the same frequency range. The abscissa value is $20 \log_{10}(w_{o}/w_{i})$, where $w_i$ and $w_o$ denote the output and input displacements of the array, respectively, recorded by point probes. 

The gates have a footprint of approximately $800 \times 50$ mm. Solutions for the single logic operations and full adder are generated in the frequency domain, using the settings outlined above. For any gate, domain probes monitor the kinetic energy density of the middle unit cells at gate inputs, which are used in conditional statements to update the output resonator elastic moduli (simulating the effect of phase change from an applied thermal stimulus) based on the probing scheme discussed at the end of Section \ref{sec2.1}. For the full adder, connections between gates are modeled using continuity boundary conditions, which transfer the displacements between boundaries. To limit the effect of reflections on solutions, Rayleigh damping is used at the input and outputs of gates (excluding connected gates). The impedance mismatch between damping layers and gate substrates is accounted for using mass and stiffness damping values that grow at a quadratic rate over the damping layer length, with the maximum values set to $\alpha=100$ s$^{-1}$ and $\beta={0.01}$ s.

The input oscillator results shown in Section \ref{sec3.2} are generated in the time domain with a time step of $t_{\text{step}} = (20f_o)^{-1}$ where $f_o=13,650$ Hz. A switching variable, $State$, sets the elastic moduli, $E_{L1}$ and $E_{L0}$, of the resonators in the upper and lower legs, respectively, using the following expressions,

\begin{subequations}
\begin{equation}
    E_{L1} = State \cdot E_a + (1-State) \cdot E_m
\end{equation}
\begin{equation}
    E_{L0} = (1-State) \cdot E_a + State \cdot E_m
\end{equation}
\end{subequations}
where $E_m$ and $E_a$ are the elastic moduli of the SMA in the martensite and austenite phases. Thus, when $State=0$, the upper leg is set to $E_m$, corresponding to an active ``$1$'' output, and the lower leg is in its bandgap state with the resonators at $E_a$. When $State=1$, the opposite is true. From the Global ODEs and DAEs module, we define a set of ODEs that model the transient effects of heating and cooling the resonators. The following ODE governs the evolution of the switching variable $State(t)$, 

\begin{equation}\label{eq3}
\frac{d\,State}{dt} = \big((1 - State) \cdot K_{\textrm{up}} \cdot L_u\big) - \big(State \cdot K_{\textrm{down}} \cdot L_{d}\big)
\end{equation}

$State$ is a dimensionless variable representing the current configuration of the metamaterial (e.g., Leg ``$1$'' or ``$0$'' active). In this ODE, the constants $K_\textrm{up}$ and $K_\textrm{down}$ are rate coefficients which control the upward and downward transition of $State$, both are set to $7,000$ s$^{-1}$. The direction of the $State$ is determined by the variables $L_{u}$ and $L_{d}$, which latch the transition of state to one direction (up or down) based on the energy probe readings. Their transition is determined by the following ODEs, 

\begin{subequations}
\begin{equation}\label{eq4}
\frac{d L_{u}}{dt} = \big(k_{l} \cdot H(\mathcal{E}_{L1} - \mathcal{E}_{L0} - \delta) \cdot (1 - L_{u})\big) - \big(k_{l} \cdot L_{u} \cdot H(State - [1-\tau])\big)
\end{equation}
\begin{equation}\label{eq5}
\frac{d L_{d}}{dt} = \big(k_{l} \cdot H(\mathcal{E}_{L0} - \mathcal{E}_{L1} - \delta) \cdot (1 - L_{d})\big) - \big(k_{l} \cdot L_{d} \cdot H(\tau-State)\big)
\end{equation}
\end{subequations}
where $L_u$ and $L_d$ are the dimensionless variables present in Eq.~\eqref{eq3} which operate to lock the evolution of $State$ in one direction until it is within a tolerance ($\tau=0.001$) of the maxima or minima of $State$ (1 or 0). Which latch is active depends on the kinetic energy density present in the two output legs denoted by the variables $\mathcal{E}_{L1}$ and $\mathcal{E}_{L0}$, which correspond to the output paths ``$1$'' and ``$0$'', respectively.

If $\mathcal{E}_{L1}$ and $\mathcal{E}_{L0}$ exceeds the threshold $\delta$ in Eq.~\eqref{eq4}, the upward latch $L_{u}$ is set and rises with a rate determined by the constant $k_l$, which is chosen to be much larger ($k_l=100,000$ s$^{-1}$) than the rate constant controlling the state, so $L_u$ will very quickly reach and lock to $1$ until $State$ reaches $1$ (within the tolerance). At this point, the upward latch is unlocked, and if the opposing condition in Eq.~\eqref{eq5}, $\mathcal{E}_{L0}-\mathcal{E}_{L1}>\delta$, is met, the downward latch is set and rises with the same rate $k_l$, thus bringing the state back down to $0$. These latches ensure that the direction of $State$ does not reverse with fluctuations in energy while the elastic modulus of the resonators is transitioning.

For the experiment, the aluminum (6061 T6) frame of the gate and nitinol resonators are manufactured using a water jet cutter. The gate has an overall footprint of $400 \times 35.2 \times 0.8$ mm. The input legs are clamped between two aluminum plates, which are then attached to an aluminum T-slot frame fixed on a vibration-isolating table. Thermal stimulation is applied using a nichrome wire heating element fixed on an insulating plate behind the metamaterial array. The on/off state of the heater is controlled by a sensor processing circuit, whose operation was elaborated on in Section~\ref{sec4.1}. Out-of-plane excitations are generated using the S233-H5FR-1107XB sealed piezoelectric bending transducers from Mide Technology. For input sensing, we utilize a $10$ mm piezoelectric disc from UXCell, which is glued to the aluminum substrate where the input paths merge. 

Displacement data is captured using a Polytech PSV-500 scanning laser Doppler vibrometer using their acquisition software running on version 9.2. The AND and OR wavefields in Figure~\ref{fig5} are frequency domain snapshots of the gates at the operating frequency of the aluminum unit cell ($f_{o,e}=6,275$ Hz). The SLDV bandwidth is set to $16$ kHz, resulting in an FFT resolution of $5$ Hz. For the input and output paths, the grid of scan points has a density of around $3$ points per centimeter. A denser grid is used to scan the unit cell array, having $5$ points per centimeter. The generator voltage is fed back to the vibrometer, so the scan of each point begins at the same point in the input cycle, ensuring phase data is accurately recorded. 

For the performance study in Section~\ref{sec4.3}, the SLDV recorded a time signal of a single point for $400$ seconds with a sample rate of $20$ kHz. In the laser software, a band-pass filter centered around the operating frequency with cutoff frequencies at $5,275$ and $7,275$ Hz was used to capture vibrations solely from applied excitations. For the computational ceiling test, time signal data is captured across the entire gate, rather than at a single point. To generate accurate data capture for each point, a pulsed excitation with triggering is used. A pulse of waves at the operating frequency is sent out for $3$ ms, then nothing for $200$ ms. Each scan is triggered to start at the beginning of the pulse, and the additional $200$ ms ensures all the waves generated from the previous pulse have dissipated, before the next point is scanned.

\bmsection*{Supporting information}

Additional supporting information may be found in the
online version of the article at the publisher’s website.

%\backmatter
\bmsection*{Author contributions}

M.N. designed and supervised the research. E.F. and M.M. conducted the theoretical analysis. E.F. carried out the numerical simulations and experimental work. E.F. and M.M. analyzed the data. The paper was written with input from all the authors.

\bmsection*{Acknowledgments}
The authors acknowledge the support of this work by the Mechanical Behavior of Materials program of the US Army Research Office (ARO), under Grant No. W911NF-23-1-0078.

\bmsection*{Conflict of interest}

The authors declare no potential conflict of interests.

\bibliography{references}

\end{document}